# An Introduction to Rule-based Modeling of Immune Receptor Signaling

John A.P. Sekar, James R. Faeder

*Department of Computational and Systems Biology, University of Pittsburgh School of Medicine, Pittsburgh, Pennsylvania.*

## 1. Introduction

It is well known that cells process external and internal signals through chemical interactions (Lodish et al., 2008; Nelson and Cox, 2013). Cells that constitute the immune system can be of many different types, such as antigen presenting cell, T-cell, B-cell, mast cell, etc. Each of these cells can have different functions, such as adaptive memory or inflammatory response, and the type and functionality of the cell is largely determined by the type and number of receptor molecules on the cell surface and the specific intracellular signaling pathways activated by those receptors (Owen et al., 2013). Given a particular biochemical signaling system in a particular immune cell, explicitly modeling and simulating kinetic interactions between molecules allows us to pose questions about system dynamics under various conditions (Aldridge et al., 2006). A model that recapitulates current experimental data can then be used to predict the results of future experiments and perturbations, and this cycle of model prediction and verification can lead to a better understanding of the system and potential clinical applications (Kitano, 2002). The promise of a mechanistic understanding has led to the enthusiastic application of chemical kinetics to biochemical signaling systems, but it has been limited by the complexity of the systems under consideration (Hlavacek et al., 2003; Borisov et al., 2005). Rule-based modeling is an approach to building and simulating chemical kinetic models that addresses this complexity (Sekar and Faeder, 2012; Chylek et al., 2013). BioNetGen (Hlavacek et al., 2006; Faeder et al., 2009), Kappa (Danos et al., 2007a, 2007b) and Simmune (Meier-Schellersheim et al., 2006; Zhang et al., 2013) are some of the more widely-used rule based frameworks. PySB provides a python framework for rule-based modeling that uses BioNetGen or Kappa to generate models (Lopez et al., 2013). In this chapter, we will explore the origins of complexity in macromolecular interactions, show how rule-based modeling can be used to address complexity, and demonstrate the construction of a model in the BioNetGen framework. Open source BioNetGen software and documentation are available at http://bionetgen.org. We highly recommend that users start with RuleBender (Smith et al., 2012), which is a graphical user interface for BioNetGen that has text highlighting and syntax checking as well as interactive visualization and simulation capabilities.

## 2. Origins of Complexity

For this chapter, we will consider a model of signaling from the FcεRI receptor present in mast cells (Faeder et al., 2003), which we will refer to as "the FcεRI model." Degranulation of mast cells and basophils plays an important role in allergic immune response, and this is initiated by signaling from FcεRI receptors on these cells. These receptors recognize the Fc portion of IgE antibodies whose Fab portion binds antigens. The activating ligand molecule is a single antigen bound to multiple antibodies, thereby providing multiple Fc sites to bind FcεRI receptors. When two or more receptors are crosslinked by a ligand, a sequence of binding and phosphorylation events on the cytoplasmic side lead to activation of Syk kinase, which subsequently promotes the degranulation response (Siraganian et al., 2010). Faeder et al. (2003) examined the dose response of Syk activation to ligand concentration and use four molecule types: a ligand, the FceRI receptor, Lyn kinase and Syk kinase.

The structural assumptions underlying the FcεRI model are summarized in Figure 1a. The model considers a bivalent ligand, i.e. one with two Fc sites for the FcεRI receptor. The receptor itself is made up of three chains: the $\alpha$ subunit capable of binding ligand, the $\beta$ subunit capable of binding Lyn and the $\gamma$ subunit capable of binding Syk. The $\beta$ and $\gamma$ subunits have phosphorylation sites present on them, and we assume that the functional states for the subunits are 'phosphorylated' and 'unphosphorylated.' On Lyn kinase, we consider two domains: a unique domain U capable of binding unphosphorylated $\beta$ subunit of receptor weakly, and an SH2 domain that binds the phosphorylated $\beta$ subunit strongly. On Syk kinase, we consider three domains: a tSH2 domain that binds phosphorylated $\gamma$ subunit of receptor, a group of phosphorylation motifs on the activation loop of the Syk kinase, and another group of phosphorylation motifs present in a linker region on Syk. Activation loop phosphorylation is known to activate the kinase activity of Syk, whereas linker phosphorylation is presumed to have a modulatory effect. Syk phosphorylated on activation loop can be considered the output of this system.

The reaction mechanisms in the model are summarized in Figure 1b. A dimer is formed by the bivalent ligand crosslinking two receptors. Lyn binds weakly using the U domain and in the dimer form, it phosphorylates the $\beta$ and $\gamma$ subunits of the adjacent receptor. The phosphorylated $\beta$ domain can bind Lyn strongly via its SH2 domain, which leads to increased Lyn-dependent phosphorylation. The phosphorylated $\gamma$ domain recruits Syk, and recruited Syk is phosphorylated in two ways: on the activation loop by Lyn recruited to the adjacent receptor, and on the linker region by Syk recruited to the adjacent receptor. The model makes conservative assumptions about how binding processes influence each other, i.e. it assumes that ligand-binding, Lyn-binding and Syk-binding events can happen independent of each other as long as the receptor they bind to has the necessary binding site exposed. Similarly, it assumes that Syk binding to receptor happens independent of its phosphorylation status. Given additional information, it is possible to modify these assumptions in the rule-based model and add context to binding events, but we will use the originally published assumptions (Faeder et al., 2003) for the purposes of this chapter.

The independence assumptions lead to many valid combinations of molecules, states and bonds that can coexist, and this leads to a large state space of chemical species, i.e. a large number of unique molecules and complexes. This phenomenon is called combinatorial complexity (Hlavacek et al., 2003) and Figure 2 shows an accounting of the possible species that can form in the FcεRI model given the interactions in the model. There are 4 variants of the free Syk molecule not bound to anything else, because of two phosphorylation motifs on Syk that can each be phosphorylated or unphosphorylated (2x2). Similarly, there are 4 variants of free receptor molecule not bound to anything else, because of two subunits that can each be phosphorylated or unphosphorylated (2x2). All 4 free receptors can bind Lyn, leading to 4 receptor-Lyn complexes. Two free receptors are $\gamma$-phosphorylated and can bind four forms of Syk each, leading to 8 receptor-Syk complexes (2x4). Similarly, two receptor-Lyn complexes have a phosphorylated $\gamma$ subunit that can bind 4 forms of Syk, leading to 8 receptor-Lyn-Syk complexes (2x4). In total there are 24 different complexes in which there is exactly one receptor and zero ligand, which we call monomeric receptor complexes without ligand. By binding a free ligand to each of these, we get 24 monomeric receptor complexes with ligand. Dimers, which are formed by a ligand crosslinking two monomeric receptor complexes, can be symmetric or asymmetric. On the symmetric dimer, the two recruited monomeric receptors are identical, and since there are 24 monomeric receptor types, there are also 24 symmetric dimers. On the asymmetric dimer, the two recruited monomeric receptors are dissimilar, and there are 276 such dissimilar pairs available (24*23/2), leading to 276 asymmetric dimer complexes. In total there are 48 monomeric receptor complexes (24+24) and 300 dimer complexes (24+276). Including free ligand, free Lyn and 4 forms of free Syk, there are 354 chemical species in the system.

## 3. Molecules in BioNetGen are structured objects that can combine to form complexes

Rule-based models can handle the combinatorics of large chemical state spaces automatically without manual curation. In the standard reaction formalism, each chemical species needs to have a unique label assigned by the modeler, but in the rule-based formalism, each chemical species is a structured graph that can be synthesized by a formal algorithm subject to the rules specified in the model. The structural building blocks for chemical species in a BioNetGen model are molecules, components of molecules, internal states of components, and bonds between pairs of components. The `molecule types` block in a BioNetGen model file specifies the types of molecules, components and internal states.

```
begin molecule types
    Lig(fc,fc)
    Rec(alpha,beta~0~P,gamma~0~P)
    Lyn(U,SH2)
    Syk(tSH2,linker~0~P,aloop~0~P)
end molecule types
```

Here, `Lig`, `Rec`, `Lyn` and `Syk` refer to the names of the types of molecules. The number and type of each molecular component is specified within brackets. The ligand `Lig` has two components named `fc` referring to binding sites for receptor. Since both sites are named identically, any attribute applicable to `fc` anywhere in the model will be equivalently applied to both sites on `Lig`. The receptor `Rec` has three components `alpha`, `beta` and `gamma`, and the `~{string}` notation denotes the internal states available to `beta` and `gamma` components, which are `~0` representing unphosphorylated and `~P` representing phosphorylated respectively. `Lyn` is defined to have two components `U` and `SH2`. `Syk` is defined to have three components `tSH2`, `linker` and `aloop`, with `linker` and `aloop` being allowed to take unphosphorylated and phosphorylated states.

All molecular variants that exist in the model are derived from permutations of internal states defined in the `molecule types` block. In this case, Lyn has one only one variant, `Lyn(U,SH2)`, whereas the receptor has four possible variants, `Rec(alpha,beta~0,gamma~0)`, `Rec(alpha,beta~0,gamma~P)`, `Rec(alpha,beta~P,gamma~0)` and `Rec(alpha,beta~P,gamma~P)` respectively. Complexes are synthesized by joining molecules using bonds between pairs of components, which is denoted using the `!` symbol followed by a bond index. Shown below are three species of increasing size:

```
Rec(alpha,beta~0,gamma~0)
Rec(alpha,beta~0!1,gamma~0).Lyn(U!1,SH2)
Rec(alpha,beta~0!1,gamma~P!2).Lyn(U!1,SH2).Syk(tSH2!2,aloop~0,linker~0)
```

The first species contains a single `Rec` molecule with nothing bound and all sites unphosphorylated (i.e. with states ~0). The second species has two molecules `Lyn` and unphosphorylated `Rec` and these are bound via their respective `U` and `beta` components. The bond label `!1` is placed adjacent to `beta` and `U` components to show that there is a binding interaction between them. The third species has three molecules `Lyn`, `Syk` and `Rec` that are bound together by a bond `!1` between `U` and `beta` components and a bond `!2` between `gamma` and `tSH2` components, with `gamma` phosphorylated as indicated by the state `~P`. Arbitrarily large complexes can be constructed in this manner and the process can be automated as discussed below.

## 4. Patterns select species with shared features

The advantage of using a structured graph specification for chemical species such as molecules and complexes is that it is possible to refer to multiple species by specifying a shared subgraph. Such a subgraph is called a **pattern** in BioNetGen and serves as a tool to partition the state space into matching versus non-matching chemical species without having to enumerate the full state space of species. Because both patterns and chemical species are graphs constructed from the same fundamental elements, one can define patterns to match any set of structural features in any arrangement of molecules within a complex, and the specified pattern may be matched by any number of molecules and complexes.

Shown below is a pattern that selects free ligand, i.e. ligand with both sites unbound.

`Lig(fc,fc)`

This pattern matches exactly one chemical species, shown below. The matches of the pattern to species are shown in red. Figure 3a shows a visualization of the match between pattern and species.

`Lig(fc,fc)`

Next, we show a pattern that selects receptors with an unbound alpha domain.

`Rec(alpha)`

Note that we have omitted the other receptor components `beta` and `gamma`, which means that their binding and modification states will not affect selection of matching species. For example, the pattern `Rec(alpha)` selects each of the following three species, visualized in Figure 3b:

```
Rec(alpha,beta~0,gamma~0)
Rec(alpha,beta~0!1,gamma~0).Lyn(U!1,SH2)
Rec(alpha,beta~0!1,gamma~P!2).Lyn(U!1,SH2).Syk(tSH2!2,aloop~0,linker~0)
```

Next, we show a pattern that selects ligand-containing complexes in which one site on the ligand is free, but the other is bound to a receptor:

`Lig(fc,fc!0).Rec(alpha!0)`

Shown below are three examples of complexes with matches shown in green, also visualized in Figure 3c.

```
Lig(fc,fc!0).Rec(alpha!0,beta~0,gamma~0)
Lig(fc,fc!0).Rec(alpha!0,beta~0!1,gamma~0).Lyn(U!1,SH2)
Lig(fc,fc!0).Rec(alpha!0,beta~0!1,gamma~P!2).Lyn(U!1,SH2).
                                    Syk(tSH2!2,aloop~0,linker~0)
```

In the full model, the patterns `Rec(alpha)` and `Lig(fc,fc!0).Rec(alpha!0)` will match all 24 monomeric receptor complexes and 24 ligand-bound monomers respectively, whose accounting was performed in Figure 2.

The BioNetGen pattern syntax enables flexible selection of different chemical species based on which structural features are specified. As discussed previously, omitting a component implies that neither its binding nor internal states will be used as match criteria. It is also possible to specify the binding state but not the internal state, e.g., `Rec(beta)` will match complexes with an unbound `beta` domain, both `beta~0` and `beta~P`. Similarly, it is also possible to specify the internal state if present, but leave the binding state unspecified (`!?`) or partially specified (`!+`). Here, `!?` matches both bound and unbound states of components, and `!?` matches all bonds with that component type irrespective of binding partner.

## 5. Reaction rules define interactions between molecules and can generate reactions

Reaction mechanisms specified in the BioNetGen language are called **reaction rules**. A reaction rule has five parts: (1) an optional label, (2) patterns specifying the properties reactants must possess to be selected by the rule, (3) an arrow indicating whether or not the rule is reversible, (4) patterns specifying the products to indicate how reactants are transformed by the rule, and (5) a set of rate laws that govern the kinetics of each reaction generated by the rule. Shown below is a reaction rule involving the three patterns specified above:

```
R1:    Lig(fc,fc) + Rec(alpha) <-> Lig(fc,fc!0).Rec(alpha!0)     kp1,km1
```

The rule is visualized in Figure 4a. Here, R1 is the name of the rule. The rule is bidirectional because of the use of the reversible arrow (<->). `Lig(fc,fc)` and `Rec(alpha)` are the reactant patterns. `Lig(fc,fc)` specifies that both sites on ligand are unbound, so it will match the free ligand species. `Rec(alpha)` matches receptors with an unbound alpha domain, so it will match all 24 monomeric receptor complexes that do not have a bound ligand. The pattern `Lig(fc,fc!0).Rec(alpha!0)` is the product of the rule. By examining it relative to the reactants, BioNetGen infers that the action of the rule is to add a bond between an available `fc` site and `alpha` subunit (highlighted in red) and can apply the rule to all combinations of matched reactant species. Since `Lig(fc,fc)` and `Rec(alpha)` match 1 and 24 species respectively, the reaction rule generates 1x24 ligand-binding reactions. Application of the rule in reverse also generates all corresponding ligand dissociation reactions, i.e. the 24 reactions that result in free ligand and a monomeric receptor complex. `kp1` and `km1` are the rate constants applicable to reactions generated from the forward and reverse directions of the rule respectively.

Unidirectional rules can be specified by using the forward arrow (->) and specifying only one rate constant. An example of a unidirectional rule is shown below, modeling transphosphorylation of receptor by recruited Lyn:

```
R4:    Lyn(U!1).Rec(beta!1,alpha!2).Lig(fc!2,fc!3).Rec(alpha!3,beta~0)-> \
       Lyn(U!1).Rec(beta!1,alpha!2).Lig(fc!2,fc!3).Rec(alpha!3,beta~P)    pLb
```

The rule named R4 is visualized in Figure 4b. Here, the reactant pattern shows four molecules connected by three bonds: a ligand bound to two receptors and `Lyn` bound to one of the receptors. The other receptor is phosphorylated on the free `beta` component i.e. transformed from ~0 on the reactant side to ~P on the product side (shown in red). The `gamma` components on the receptors are omitted since they do not explicitly affect this process. `pLb` is a parameter that specifies the first order phosphorylation rate constant.

Because patterns used in rules can get large, it helps to sort the specified structures into **reaction center** and **reaction context** to aid their comprehension and discussion. The reaction center is the set of structures that are modified by the rule and thereby defines the action of the rule. The reaction center of rule R1 specified above is the set of binding sites `fc` and `alpha` on the reactant side, and the `fc-alpha` bond on the product side. The reaction center of rule R4 is the internal state of the `beta` component that was modified from ~0 to ~P. The reaction context includes the remaining structures that are specified in the rule but are not modified by the rule. This establishes the set of local conditions under which the action of the rule is allowed to happen. For rule R1, the reaction context is the presence of a second unbound `fc` site. For rule R4, the reaction context is the unbound status of the `beta` site that was phosphorylated, and the set of three bonds that constitute the Lyn-recruited-to-dimer configuration. Given a rule, the reaction center is identified by the differences between the reactants and products of the rule and the reaction context is identified by structures preserved on both reactants and products.

Rules interact with each other when the reaction center of one rule overlaps with the reaction context of another. For example, the `fc-alpha` bond that is formed in rule R1 is necessary for the Lyn-recruited-to-dimer configuration that is context for rule R4, implying that R1 enables R4. These interactions constitute the flow of signal through the system and reaction rules enable reconstituting a signaling pathway from the fundamental kinetic processes. The FcεRI model used in this chapter has 19 reaction rules modeling free ligand binding (1), receptor crosslinking by ligand (1), Lyn binding to receptor (2), Syk binding to receptor (1), transphosphorylation by Lyn on receptor and Syk sites (6), transphosphorylation by Syk on Syk (2) and background dephosphorylation processes (6) respectively. In the BioNetGen model file, rules are listed in the `reaction rules` block:

```
begin reaction rules
# Ligand-receptor binding
R1: Rec(alpha) + Lig(fc,fc) <-> Rec(alpha!1).Lig(fc!1,fc)  kp1, km1

# Receptor-aggregation
R2: Rec(alpha) + Lig(fc,fc!0).Rec(alpha!0) <-> \
Rec(alpha!1).Lig(fc!1,fc!0).Rec(alpha!0)  kp2, km2

# Constitutive Lyn-receptor binding
R3: Rec(beta~0) + Lyn(U,SH2) <-> Rec(beta~0!1).Lyn(U!1,SH2)  kpL, kmL

# Transphosphorylation of beta by constitutive Lyn
R4: Lyn(U!1).Rec(beta!1,alpha!2).Lig(fc!2,fc!3).Rec(alpha!3,beta~0)->\
Lyn(U!1).Rec(beta!1,alpha!2).Lig(fc!2,fc!3).Rec(alpha!3,beta~P) pLb
...
...
end reaction rules
```

Text beginning with # and ending on a line break can be used to provide useful comments and annotations anywhere in the model, and they are ignored by the BioNetGen processor. The character \ can be used to split rules across lines when the rules get really large. Within patterns, the order of components between brackets does not matter. Similarly, the specific bond index used for bonds does not matter, as long as the intended pair of components have the same unique bond index. BioNetGen also allows more complex specifications for the rate law that are out of the scope of this chapter (Chylek et al., 2015).

## 6. Observables and functions define the outputs of the model

There are two types of outputs in the BioNetGen model: measurements of species concentrations called observables and functions of those observables. In a rule-based model, pattern matching can be exploited to generate sums of species automatically. Observables are of two types: counting unique complexes only, or counting individual matches of patterns within complexes. These are specified using the `Species` and `Molecule` keywords respectively within the observables block. For the FcεRI model, we are interested in the partitioning of individual receptor and Syk molecules into various configurations, so we will only use `Molecule` observables. The observables block used for this model is:

```
begin observables
    Molecules   Dimer       Rec(alpha!1).Lig(fc!1,fc!2).Rec(alpha!2)
    Molecules   TotalRec    Rec()
    Molecules   ActiveSyk   Syk(aloop~P)
    Molecules   TotalSyk    Syk()
    Molecules   BetaP       Rec(beta~P!?)
    Molecules   GammaP      Rec(gamma~P!?)
end observables
```

Here, the pattern `Rec(alpha!1).Lig(fc!1,fc!2).Rec(alpha!2)` counts all ligand-crosslinked dimer species, of which there are 300 types in the model. Since it is a `Molecule` observable and the pattern will match twice into each dimer, each dimer will be counted twice and the resultant sum will be the total number of receptors in dimers. The pattern `Syk(aloop~P)` counts number of Syk molecules with phosphorylated activation loop, and the patterns `Rec(beta~P!?)` and `Rec(gamma~P!?)` count receptor molecules that are respectively phosphorylated on `beta` and `gamma` respectively. Here `!?` was used to ensure that the observable counted both unbound and bound forms of `beta` and `gamma` components. This was not necessary for `aloop` since there are no binding rules specified for `aloop`. Also, the two observables for $\beta$ and $\gamma$ phosphorylation will have some species that are counted for both observables, since receptors can be phosphorylated on both `beta` and `gamma` sites. Finally, observables `Rec()` and `Syk()` simply count all receptor molecules and Syk molecules in the system respectively, since no component-matching terms are specified.

Functions of observables can be defined in the `functions` block. For the FcεRI model, we will consider three outputs: fraction of receptor in dimers, fraction of Syk that is active (i.e. phosphorylated on the activation loop), and ratio of phosphorylated gamma and beta sites on receptors.

```
begin functions
    DimerFraction                       Dimer/TotalRec
    AutoPhosphorylatedSykFraction       ActiveSyk/TotalSyk
    GammaBetaPhosphorylationRatio       GammaP/BetaP
end functions
```

# 7. Parameterizing a model requires geometric terms, rate constants and species concentrations

Parameterizing a BioNetGen model requires a self-consistent set of units for concentrations and rate constants. For this particular model, we have only used unimolecular and bimolecular reactions and we would like to measure protein fluxes in molecules per second. We will limit ourselves to considering a single cell surrounded by a milieu of freely diffusing ligand. We will also specify units as comments using # symbol in order to be clear to the modeler. These comments will be ignored by BioNetGen.

First, we will need to specify fundamental constants such as $\pi$ and Avogadro's number $N_A$ to be used in our calculations.

```
NA          6.023e23    # units: molecules/mol
pi          3.14159     # no unit
```

Next, we will need to specify concentrations of proteins and ligands that reflect the experimental setup. Ligand concentration for this system is typically provided in nanomolar values and has to be converted to molecule numbers. One molar concentration is one mol/liter, so the conversion is achieved by multiplying molar concentration with Avogadro constant and volume of the external milieu, which in turn is estimated from a cell density of million cells per milliliter. A conversion factor of 1e3 is required to convert milliliter volumes to liters.

```
celldensity     1e6*1e3             # units: cells/L
vol_ext         1/celldensity       # units: L/cell*(1 cell) = L
Lig_conc        1e-9                # units: molar
Lig_tot         Lig_conc*vol_ext*NA # units: molecules
```

Receptor numbers are usually provided as molecule numbers per cell, so they are used directly. Lyn and Syk numbers are estimated relative to receptor numbers, so they are specified using proportional factors.

```
Rec_tot         4e5              # units: molecules
fLyn            0.07
fSyk            1
Lyn_tot         fLyn*Rec_tot
Syk_tot         fSyk*Rec_tot
```

Unimolecular rate constants such as those for phosphorylation, dephosphorylation and bond dissociation are usually known or estimated in per second units, so they can be used directly. Here, we show one example of each: ligand dissociation constant (km1), rate of $\beta$-phosphorylation by Lyn (pLb) and rate of background dephosphorylation at the membrane (dm).

```
km1     0.01    # units: /s
pLb     30      # units: /s
dm      20      # units: /s
```

Bimolecular association rate constants are usually provided in units of per molar per second. Converting to per molecule per second requires dividing by Avogadro constant and the respective volume in which the bimolecular association occurs. We show three examples below: association rate constant for free ligand which needs to be scaled by external volume (kp1), association rate constant for ligand crosslinking which occurs in the membrane (kp2), and association rate constant for Lyn binding to unphosphorylated β site which occurs in the cytoplasm (`kpL`). For a simple BioNetGen model, scaling of association rate constants can be performed manually, but for more complicated systems it is recommended to use the compartmental extension to the BioNetGen framework (Harris et al., 2009a) where such scaling is performed automatically.

```
kp1    1e7/(NA*vol_ext)        # units: /molecule/s
kp2    1e6/(NA*vol_mem)        # units: /molecule/s
kpL    4.2e7/(NA*vol_cell)     # units: /molecule/s
```

External volume was calculated previously from cell density. Volume of the membrane is specified by multiplying surface area of the cell with an effective width of 10 nanometers (Harris et al., 2009a). Volume and surface area of the cell are both estimated by assuming a spherical cell with 7 micron radius. Cubic meter volume is converted to liter units by multiplying by 1e3.

```
rad_cell    7e-6                        # units: meter
vol_cell    1e3*(4/3)*pi*rad_cell^3     # units: L
surf_area   4*pi*rad_cell^2             # units: squared meter
eff_width   10e-9                       # units: meter
vol_mem     1e3*surf_area*eff_width     # units: L
```

Here, the association constant for membrane reactions (`kp2`) was specified as a three-dimensional association rate constant (units: per molar per second, i.e. per "mol per liter" per second), which necessitates a conversion of the two-dimensional surface area of the membrane to a three-dimensional volume using an effective width term. An alternate parameterization would be to use a two dimensional association rate constant (units: per "mol per squared meter" per second), in which case it is sufficient to multiply by surface area only and not the effective width.

The parameters and parameter expressions are specified in the `parameters` block in the model file.

```
begin parameters
    NA              6.023e23                # units: molecules/mol
    pi              3.14159
    rad_cell        7e-6                    # units: meter
    vol_cell        1e3*(4/3)*pi*rad_cell^3 # units: L
    celldensity     1e6*1e3                 # units: cells/L
    vol_ext         1/celldensity           # units: L/cell*(1 cell)= L
    Lig_conc        1e-9                    # units: molar
    Lig_tot         Lig_conc*vol_ext*NA     # units: molecules
    ...
    ...
end parameters
```

Initial concentrations for chemical species are specified in the `seed species` block. The concentrations can be specified in terms of parameter expressions defined in the `parameters` block.

```
begin seed species
    Lig(fc,fc)                  Lig_tot
    Lyn(U,SH2)                  Lyn_tot
    Syk(tSH2,linker~0,aloop~0)  Syk_tot
    Rec(alpha,beta~0,gamma~0)   Rec_tot
end seed species
```

In the seed species block, only fully defined molecules and complexes must be specified. In other words, every molecule should have every component present, and every component should have internal and binding states clearly specified. For this model, the unbound and unphosphorylated forms of all molecules are considered to be the seed species. Depending on how the system is simulated the remaining species that can occur in the system will either be generated by iterative application of the rules to the seed species by a process called network generation, or they will be discovered "on-the-fly" as the system is simulated and rules modify the state of the initial species (Faeder et al., 2009).

## 8. Model actions are used to simulate and analyze a model

The complete model specification has blocks for parameters, molecule types, reaction rules, observables and seed species, and is stored in a text file with the extension ".bngl". Model actions are appended to the end of the model file in order to call tools that use the model specification. This enables the specified model to be converted into a reaction network, simulated using multiple available methods, exported to other frameworks, visualized at different resolutions, and also scanned over parameter ranges to reveal parameter sensitivity and bifurcation properties. A comprehensive actions and arguments guide is available at http://bionetgen.org. Here, we will discuss the use of three model actions: `generate_network()`, `simulate()` and `parameter_scan()`.

### 8.1. A reaction network can be generated from a rule-based model

The following action is used to generate a reaction network whose kinetics are equivalent to the specified rule-based model:

`generate_network({overwrite=>1})`

This command iteratively applies the reaction rules to the seed species in order to generate new configurations of molecules and complexes and new reactions (Faeder et al., 2009). The full state space of molecules, complexes and reactions is written to a file with the extension ".net". The overwrite command is to ensure that when network generation is performed a second time, any past generated networks of the same model are overwritten. For the full model with 4 seed species and 19 rules, the `generate_network()` action results in 354 total number of species and 3680 reactions (Faeder et al., 2003).

One of the features of network generation is that the calculation of statistical factors for reactions is handled automatically. For this reason, it is advantageous to build rule-based models even for small reaction networks. We will show using simple examples how BioNetGen handles symmetry and multiplicity.

Multiplicity factors are needed when the same rule can be applied to multiple identical parts of the same reactant species, which increases the effective rate by a factor. For example, consider the molecule types `A(b)` and `B(a,a)` and the reaction rule:

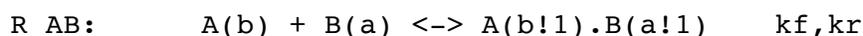
`R_AB:     A(b) + B(a) <-> A(b!1).B(a!1)    kf,kr`

Now consider how to parameterize two pairs of reversible binding reactions generated from the rule: binding in which both sites on B are free, and binding in which one of the sites is already occupied.

```
A(b)   + B(a,a)              <-> A(b!1).B(a!1,a)
A(b)   + B(a,a!1).A(a!1) <-> A(b!2).B(a!2,a!1).A(b!1)
```

The pattern `B(a)` matches to species `B(a,a)` in two different ways, i.e. using one or the other free binding site. This means that the effective rate of the reaction is double what one would expect if there was only one free site on B. BioNetGen would detect this and calculate a multiplicity factor of 2 for the forward reaction, i.e.

```
A(b)   + B(a,a)              <-> A(b!1).B(a!1,a)                    2*kf,kr
```

For the second pair of reactions, the product pattern `A(b!1).B(a!1)` would match twice into the species `A(b!2).B(a!2,a!1).A(b!1)` since there are two identical bonds that could be matched. In this case, it is the reverse reaction that needs to be multiplied by two, since the presence of two identical bonds would double the effective rate of bond breaking, i.e.

```
A(b)   + B(a,a!1).A(a!1) <-> A(b!2).B(a!2,a!1).A(b!1)   kf,2*kr
```

Symmetry factors are needed when the action of a rule modifies at least two instances of a species simultaneously and identically. For example, consider the molecule type `A(b~0~1)` and the reaction rule:

```
R_AA:      A(b) + A(b) -> A(b!1).A(b!1)        kf
```

The reaction rule would generate three reactions:

```
A(b~0) + A(b~1) -> A(b~0!1).A(b~1!1)
A(b~0) + A(b~0) -> A(b~0!1).A(b~0!1)
A(b~1) + A(b~1) -> A(b~1!1).A(b~1!1)
```

The first reaction involves two different reactant species, `A(b~0)` and `A(b~1)` respectively. If $N_0$ and $N_1$ are the concentrations of `A(b~0)` and `A(b~1)` respectively, then the first reaction would have a rate of $k_f N_0 N_1$.

The second reaction operates on two instances of the same species `A(b~0)`, and the rate is determined to be $k_f \frac{N_0(N_0-1)}{2}$, where $\frac{N_0(N_0-1)}{2}$ is the number of ways to select two molecules out of $N_0$ molecules. For high $N_0$, $k_f \frac{N_0(N_0-1)}{2} \approx k_f \frac{N_0^2}{2}$, but a rate calculation similar to the first reaction would only yield $k_f N_0^2$. Therefore, a factor of ½ needs to be applied for the second reaction and not the first. Similarly, the third reaction would also have an effective rate of $k_f \frac{N_1^2}{2}$ since it operates on two instances of the same species `A(b~1)`. Using graph isomorphism calculations (Hogg et al., 2014), BioNetGen estimates the correct symmetric factor adjustment for the rate constants of all generated reactions, i.e.

```
A(b~0) + A(b~1) -> A(b~0!1).A(b~1!1)   kf
A(b~0) + A(b~0) -> A(b~0!1).A(b~0!1)   0.5*kf
A(b~1) + A(b~1) -> A(b~1!1).A(b~1!1)   0.5*kf
```

## 8.2. Model simulation and parameter scans reveal insight about model behavior

The specified model is simulated using the `simulate()` action.

```
simulate({method=>"ode",t_end=>600,n_steps=>10,print_functions=>1})
```

There are three simulation methods that can be used: ODE integration (`method=>"ode"`), Gillespie's stochastic simulation algorithm (`method=>"ssa"`) (Gillespie, 1977) and network-free stochastic simulation (`method=>"nf"`). The ODE and SSA methods require the reaction network to be generated prior to simulation. Network-free simulation uses the NFsim algorithm (Sneddon et al., 2011), which was designed to exactly simulate kinetics without having to generate the full reaction network in advance of the simulation. It generates results indistinguishable from SSA and is useful for models that generate disproportionately large or infinite reaction networks, e.g. models with oligomerization rules. In these networks, the number of unique species may be much larger than the actual number of molecules and complexes present during simulation and only a fraction of the state space may be populated at any given time. In practical terms, the network-free simulation method typically leads to faster simulations than the corresponding SSA when the number of possible species is more than several hundred to a thousand (Sneddon et al., 2011).

The outputs of the `simulate()` action are trajectory files with extensions ".cdat" and ".gdat". The ".cdat" file is generated by ODE and SSA methods and contains concentrations of all molecules and complexes over time. The ".gdat" file is generated by all three methods and contains the measured values over time for the observables defined in the `observables` block. If the `print_functions=>1` flag is used, then the ".gdat" files include columns for each of the specified functions in the `functions` block also. The trajectory files can be imported into any data-processing software such as Microsoft Excel, OpenOffice Calc or MatLab.

BioNetGen provides additional actions in order to run multiple simulations of the same model at different parameter settings, such as `setParameter()`, `setConcentrations()`, `resetConcentrations()`, etc. It also provides a command to run batch simulations called `parameter_scan()`, which enables scanning over a particular parameter given a range of values while keeping other parameters constant. `parameter_scan()` takes the same arguments as `simulate()` shown above, and additionally takes arguments that specify the range of values to be scanned over and the reset behavior between simulations. The results of `parameter_scan()` are stored in a similar form to ".gdat" and ".cdat" files, but with the extension ".scan". Here, we show how parameter scans can be used to elucidate aspects of model behavior.

Shown in Figure 5 are results from three parameter scans for each of the three outputs defined in the `functions` block, i.e. receptor fraction in dimers, ratio of $\gamma/\beta$ phosphorylation and active Syk fraction. The parameter scan over ligand concentration was performed using the command:

```
parameter_scan({method=>"ode",parameter=>"Lig_conc",par_min=>1e-12, \
par_max=>1e-,n_scan_pts=>50, log_scale=>1,reset_conc=>1, \
print_functions=>1,t_end=>600,n_steps=>5})
```

Note that `method`, `print_functions`, `t_end` and `n_steps` arguments are used the same as in the `simulate()` command. Additionally, `parameter` specifies the parameter to scan over, `par_min` and `par_max` establish the minimum and maximum of the range of values to be used, `n_scan_pts` determines how many points to sample within the range. `log_scale=>1` sets the spacing between points to be logarithmic, which enables scanning over many orders of magnitude. For smaller ranges, the default setting of `log_scale=>0` is used, which ensures that the sampled points are spaced equally. `reset_conc=>1` ensures that when a new simulation is run, all parameter values and species concentrations are reset instead of being carried over from the previous simulation. In the

`parameter_scan()` command shown above, we enable scanning over ligand concentration from 1e-12 M to 1e-3 M. The other two `parameter_scan()` methods shown below scan over Lyn/receptor ratio and Syk/receptor ratio between the range of 0.01 to 10 and use the same remaining arguments as the first one. Action statements can also be split across multiple lines using the \ character.

```
parameter_scan({parameter=>"fLyn",par_min=>0.01,par_max=>10,\
n_scan_pts=>50,log_scale=>1,...})
parameter_scan({parameter=>"fSyk",par_min=>0.01,par_max=>10,\
n_scan_pts=>50,log_scale=>1,...})
```

In systems such as the FcεRI model where crosslinking by a ligand or scaffold is necessary for signaling, one can observe a phenomenon called high dose inhibition. As seen in Figure 5, at high concentrations of ligand, the fraction of receptors in dimers actually decreases to negligible values, and there is a corresponding decrease in the active Syk fraction. This phenomenon occurs because binding of free ligand competes with the crosslinking mechanism for free monomeric receptors, and free ligand binding dominates at high ligand concentrations, leading to decreased crosslinking and fewer dimers (Nag et al., 2010). Another mechanistic insight that can be derived for this system is the modulation of $\gamma/\beta$ phosphorylation by Lyn and Syk numbers. In the model, $\beta$ sites are phosphorylated by Lyn, as well as protected by Lyn-binding. On the other hand, $\gamma$ sites are phosphorylated by Lyn, but protected by Syk-binding. As Lyn numbers increase, the ratio of $\gamma/\beta$ phosphorylation decreases, since more $\beta$ sites are being protected by Lyn-binding (Faeder et al., 2003). Similarly, as Syk numbers increase, the ratio of $\gamma/\beta$ phosphorylation increases, since more $\gamma$ sites are being protected by Syk-binding (Faeder et al., 2003).

## 9. Advanced Methods

In the previous section, we covered elementary actions that can be applied to BioNetGen models, such as network generation and simulation by ODE, SSA (Gillespie, 1977) or NFsim (Sneddon et al., 2011). In addition to the basic specification demonstrated in this chapter, extensions to the BioNetGen language include specification of compartments and transport rules (Harris et al., 2009a), use of observables and local functions in rate laws (Sneddon et al., 2011) and preservation of detailed balance (Hogg, 2013). In addition to these, a number of improvements to simulation algorithms have been developed that are centered on the BioNetGen specification. For models where network-free simulation can occupy too much memory due to large numbers of certain molecules and complexes, a hybrid approach between network-based and network-free methods can be used (Hogg et al., 2014). The efficiency of each SSA simulation can be improved by enabling tau-leaping procedures (Harris et al., 2009b), and the sampling of rare stochastic events can be improved by weighted ensemble simulation (Donovan et al., 2013). A comprehensive BioNetGen actions and arguments guide is available at http://bionetgen.org.

A number of external software have interfaces to BioNetGen or utilize BioNetGen as a back end. These include modeling environments that facilitate model construction using programmatic, tabular or graphical interfaces, such as VCell (Moraru et al., 2008), pySB (Lopez et al., 2013), rxncon (Tiger et al., 2012) and BioUML (Kolpakov et al., 2006). Parameter sensitivity analysis and fitting to experimental data can be performed for BioNetGen models using the ptempest toolbox for MatLab (Hogg, 2013) and the BioNetFit software (Thomas et al., 2015). Recently, we have developed advanced visualization tools for showcasing individual rules and interactions between rules (Sekar et al., n.d.), ported BioNetGen model specification for the MCell simulator that explicitly simulates spatial diffusion and reactions (Stefan et al., 2014), and enabled conversion of reaction networks to rule-based specifications (Tapia and Faeder, 2013).

Kappa (http://kappalanguage.org/) and Simmune (http://simmune.org) are other rule-based modeling frameworks with similar rule-based specifications to BioNetGen, and they have their own constellations of related tools and software. A common interchange format SBML-Multi is being developed to enable models to be translated and analyzed across rule-based frameworks (http://sbml.org)

# 10. Resources

The BioNetWiki (http://bionetgen.org) is the main resource for all BioNetGen related tools. On the Downloads page, links are provided for the standalone BioNetGen distribution as well as the distribution bundled with RuleBender, a graphical user interface. Currently, precompiled packages are distributed for Windows, OSX and Linux platforms. On the Tutorials page, one can find models that have been presented or published as parts of tutorials, including the model used in this chapter. For a comprehensive reference of BioNetGen syntax and usage, see Faeder, Blinov, and Hlavacek (2009). An updated online version is maintained on the wiki, with additional documentation on model visualization, bifurcation analysis, tau leaping, SBML import and other advanced tools. Also on the wiki is a comprehensive reference for model actions and arguments used in BioNetGen. Sekar and Faeder (2012) provide a more detailed tutorial on BioNetGen that includes compartmental specification and the reconstruction of a large signaling pathway. Chylek et al. (2014) provide a current review of rule-based methods and Chylek et al. (2015) give examples that illustrate a number of advanced modeling features. The formalisms used in BioNetGen are explained in Blinov et al. (2006) and Hogg et al. (2014). For more detailed examples of rule-based models of immunoreceptor signaling, see the libraries of rules for FcεRI (Chylek et al., 2014b) and the large scale model of T cell receptor signaling (Chylek et al., 2014a), as well as Chapter 14 in this volume.

**Figure Legends**

**Figure 1: Overview of the FcεRI model. (A) Contact Map.** This diagram summarizes the molecules, sites and binding interactions in the system. The ligand has two identical sites named `fc` that bind the receptor. The receptor has `alpha`, `beta` and `gamma` subunits, each modeled as a site of the receptor. The `alpha` subunit is used to bind the ligand, the beta subunit to bind Lyn kinase, and gamma subunit to bind Syk kinase. `beta` and `gamma` subunit have phosphorylated (`P`) and unphosphorylated (`0`) states available to them. Lyn can bind receptor either through its `U` domain or its `SH2` domain. Syk binds receptor through its `tSH2` domain. Syk has two groups of phosphorylation sites, titled `aloop` and `linker` respectively which take unphosphorylated and phosphorylated states. **(B) Transphosphorylation mechanisms.** Receptors can be crosslinked to form dimers by the bivalent ligand. Lyn bound on one receptor can phosphorylate beta and gamma subunits on the adjacent (trans) receptor. Similarly, Syk bound to one receptor is phosphorylated by Lyn and Syk bound to the other.

**Figure 2: Accounting for Molecules and Complexes.** In the model, there is one type of free ligand molecule, and one type of free Lyn molecule respectively. Because sites can be independently phosphorylated, there are four types of free Syk molecules and four types of free receptor molecules respectively. Lyn binding to free receptor gives rise to four Rec-Lyn complexes. Syk binding to free receptor leads to eight Rec-Syk complexes. Syk binding to Rec-Lyn leads to eight Rec-Lyn-Syk complexes. In total, there are 24 complexes that have one receptor to which no ligand is bound, called 24 monomeric receptor complexes. When free ligand binds, this leads to 24 ligand-bound monomers. On crosslinked dimers, symmetry between the two recruited monomeric receptors leads to 24 symmetric dimers. When the recruited monomeric receptors are asymmetric, there are 276 combinations that are possible, leading to 276 asymmetric dimers.

**Figure 3: Pattern Matching to Complexes.** Shown are three patterns and matching of those patterns to subgraphs within chemical species (molecules and complexes). The specific matches are shown in color and the unmatched parts are grayed out. **(A)** Pattern `Lig(fc,fc)` matching the one free ligand species (red). **(B)** Pattern `Rec(alpha)` matching three complexes with an unbound alpha domain (blue). **(C)** Pattern `Lig(fc,fc!0).Rec(alpha!0)` matching three complexes in which one site of the ligand is occupied by receptor (green). In the FcεRI model, `Rec(alpha)` matches all 24 monomeric receptor complexes and `Lig(fc,fc!0).Rec(alpha!0)` matches all 24 ligand-bound monomers.

**Figure 4: Reaction Rules.** In a reaction rule, only the parts necessary for a kinetic process need to be specified. **(A) Reversible Binding.** Reaction rule R1 models reversible binding of free ligand and receptor. On the receptor pattern, only alpha subunit is specified because of the assumed independence between ligand-binding and other binding processes. **(B) Transphosphorylation Rule**. Reaction rule R4 models phosphorylation of `beta` domain by Lyn recruited to the adjacent (trans) receptor in a dimer. The pattern used indicates the minimum conditions for this process: a cross-linked dimer, an unphosphorylated beta domain substrate on side of the dimer, and Lyn bound to beta domain on the other side.

**Figure 5: Simulation Results.** Simulated dose-response curves from three different parameter scans (rows) of the model. For each scan, the outputs measured were fraction of receptors in dimers, ratio of phosphorylation between $\gamma$ and $\beta$ sites, and fraction of Syk that is active (columns). Varying ligand concentration showed that dimerization and Syk activation are inhibited at high ligand doses. Varying Lyn/receptor ratio showed that high concentrations of Lyn protect $\beta$ sites from dephosphorylation.

Varying Syk/receptor ratio showed that high concentrations of Syk protect $\gamma$ sites from dephosphorylation.

Figure 1

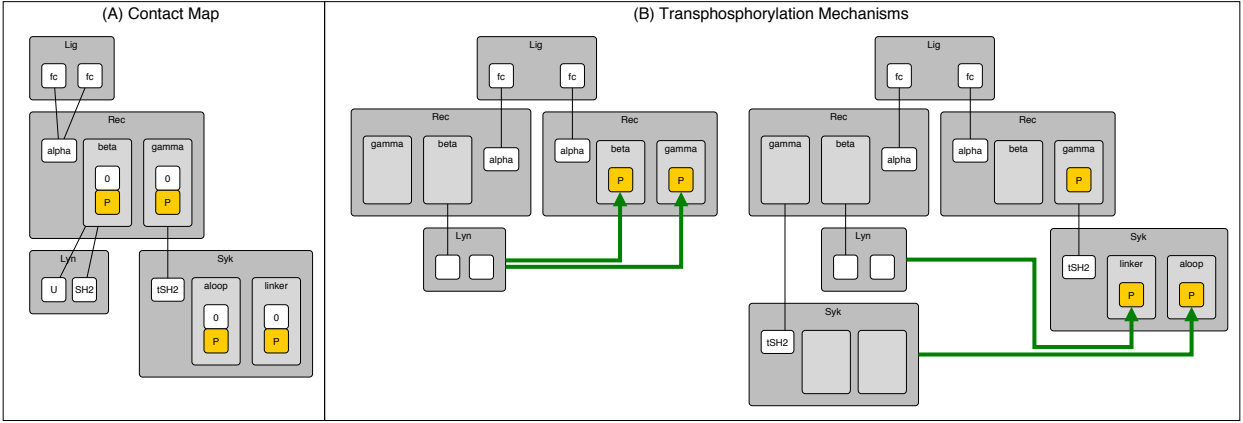

Figure 2

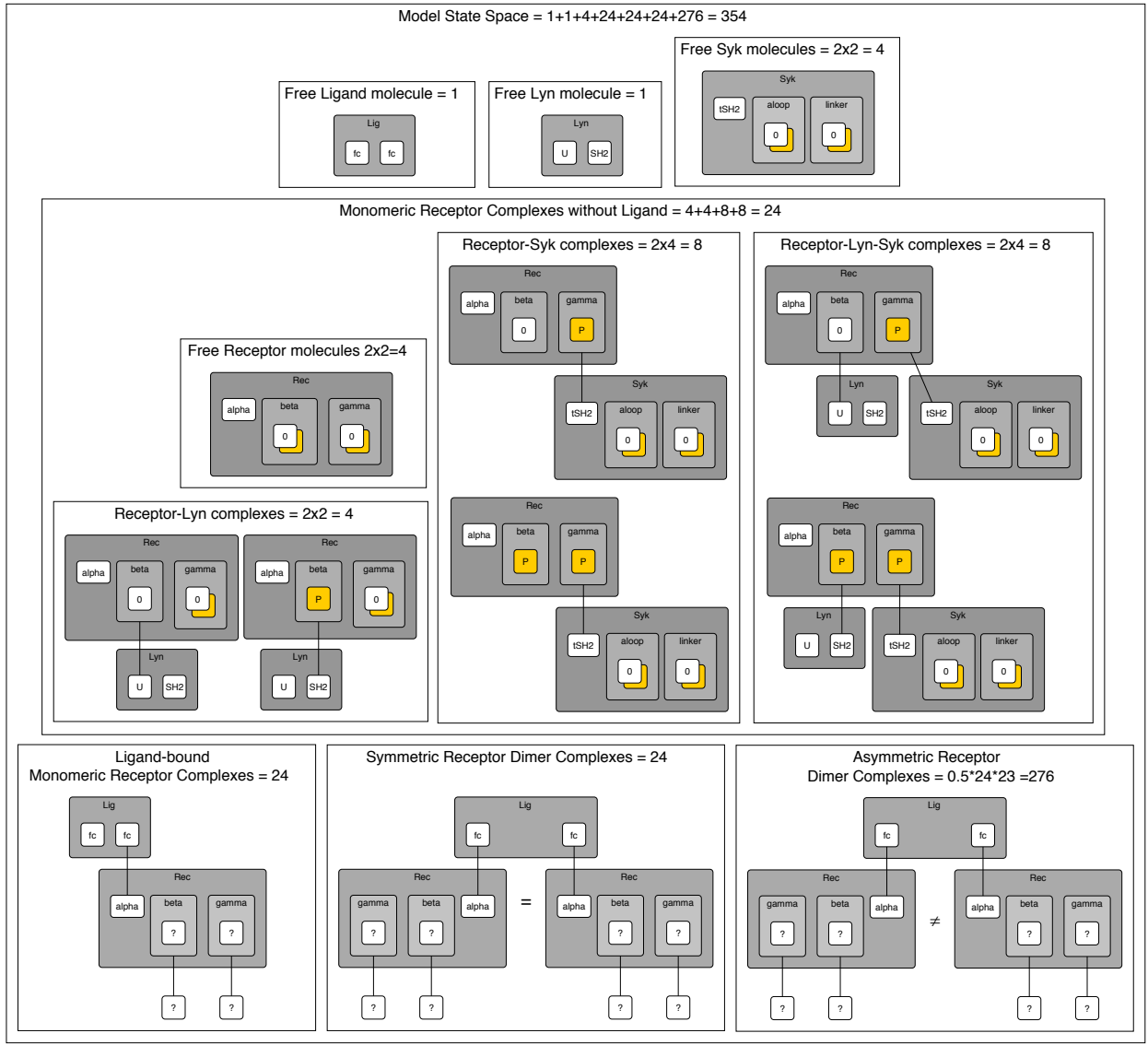

Figure 3

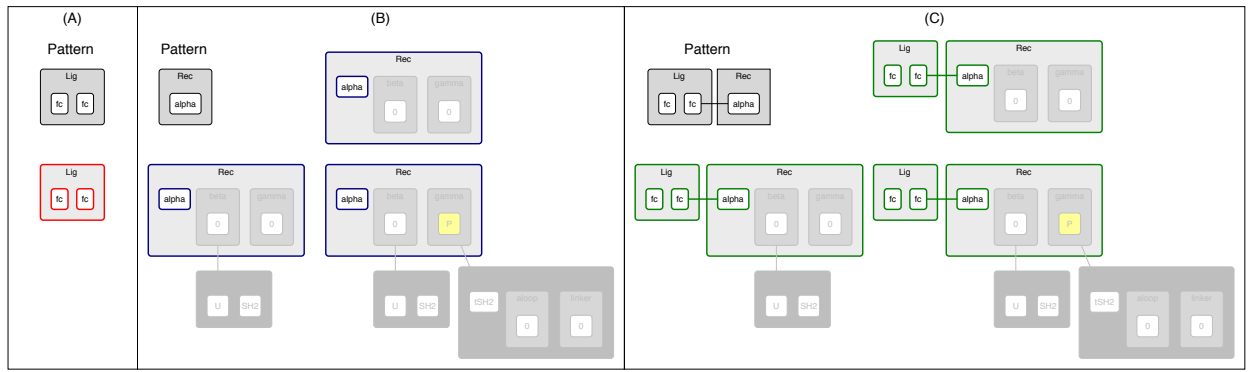

Figure 4

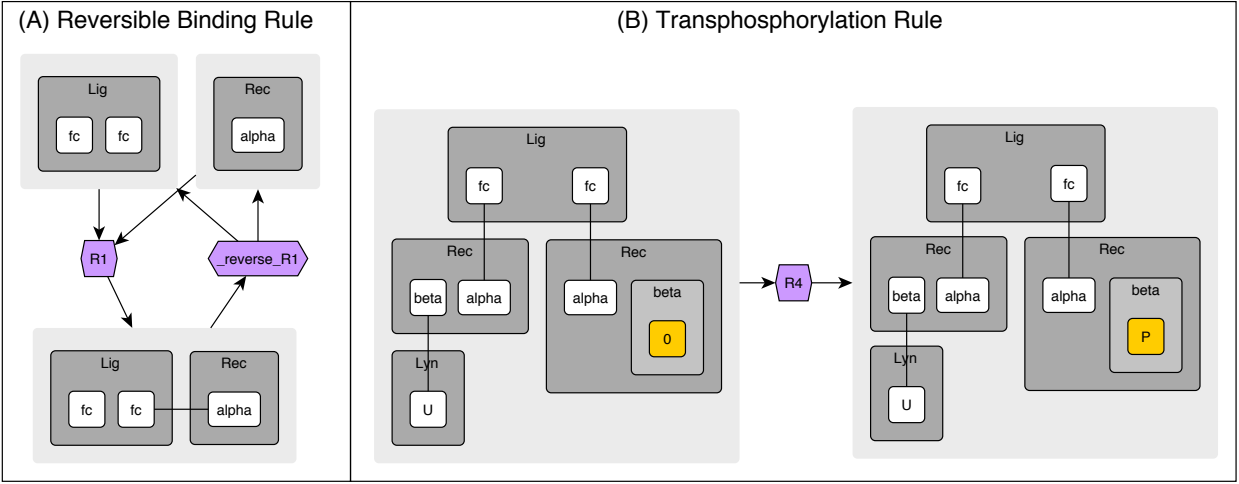

Figure 5

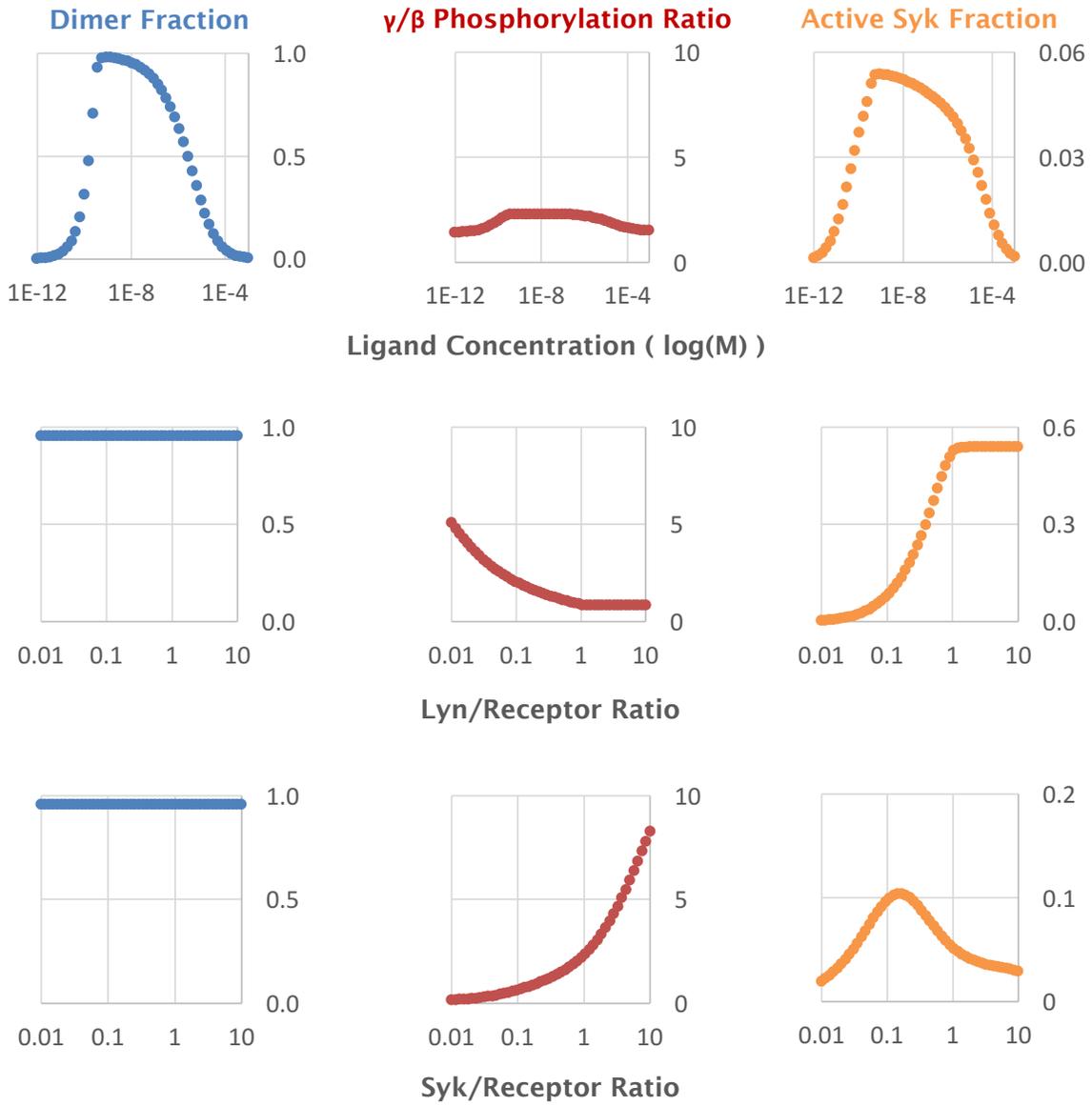